\documentclass[prl,10pt,notitlepage,twocolumn,superscriptaddress]{revtex4}

\usepackage{graphicx}
\usepackage{amsmath}
\usepackage{amssymb}
\usepackage{amsfonts}
\usepackage{color}
\usepackage{xspace}
\usepackage{subfigure}

\newcommand{\moly}{Mo$_3$S$_7$(dmit)$_3$\xspace}

\newcommand{\ket}[1]{| #1 \rangle}

\begin{document}

\title{Emergence of quasi-one-dimensional physics in {\moly}, \\ a nearly-isotropic three-dimensional molecular crystal}

\author{A. C. Jacko}\affiliation{School of Mathematics and Physics, The University of Queensland, Brisbane, Queensland, 4072, Australia}
\author{C. Janani}\affiliation{School of Mathematics and Physics, The University of Queensland, Brisbane, Queensland, 4072, Australia}
\author{Klaus Koepernik}\affiliation{IFW Dresden e.V., PO Box 270116, D-01171 Dresden, Germany}
\author{B. J. Powell}\affiliation{School of Mathematics and Physics, The University of Queensland, Brisbane, Queensland, 4072, Australia}

\begin{abstract}
We report density functional theory calculations for Mo$_3$S$_7$(dmit)$_3$. We derive an ab initio tight-binding model from overlaps of Wannier orbitals; finding a layered model with interlayer hopping terms $\sim3/4$ the size of the in-plane terms. The in-plane Hamiltonian interpolates the kagom\'e and honeycomb lattices. It supports states localized to dodecahedral rings within the plane, which populate  one-dimensional (1D) bands and lead to a quasi-1D spin-one model on a layered honeycomb lattice once interactions are included. Two lines of Dirac cones also cross the Fermi energy.
\end{abstract}

\maketitle

Understanding emergence, the observation that systems can have fundamentally different properties from their constituent parts, is one of the most important challenges in modern physics \cite{anderson72,dagotto05,levin05,schmidt09,schultz12,chakhalian14}. 
For example, spin liquids may realise  emergent gauge fields and deconfined fractionalised excitations \cite{balents10}.
However, predicting emergent behaviour in condensed matter systems is incredibly challenging. Thus, identifying motiefs that lead to certain emergent properties is an extremely valuable step towards the goal of rational design of complex materials.

Molecular crystals are an exciting test ground for finding and understanding new emergent states of matter. 
They often display competition between multilple emergent strongly correlated ground states \cite{powell11, dressel07}. This, and the flexibility of organic chemistry, means that the emergent physics is often tunable by subtle chemical and physical modification \cite{taniguchi99,dressel07,seo15}.

\begin{figure}
\begin{center}
\includegraphics[width=\columnwidth]{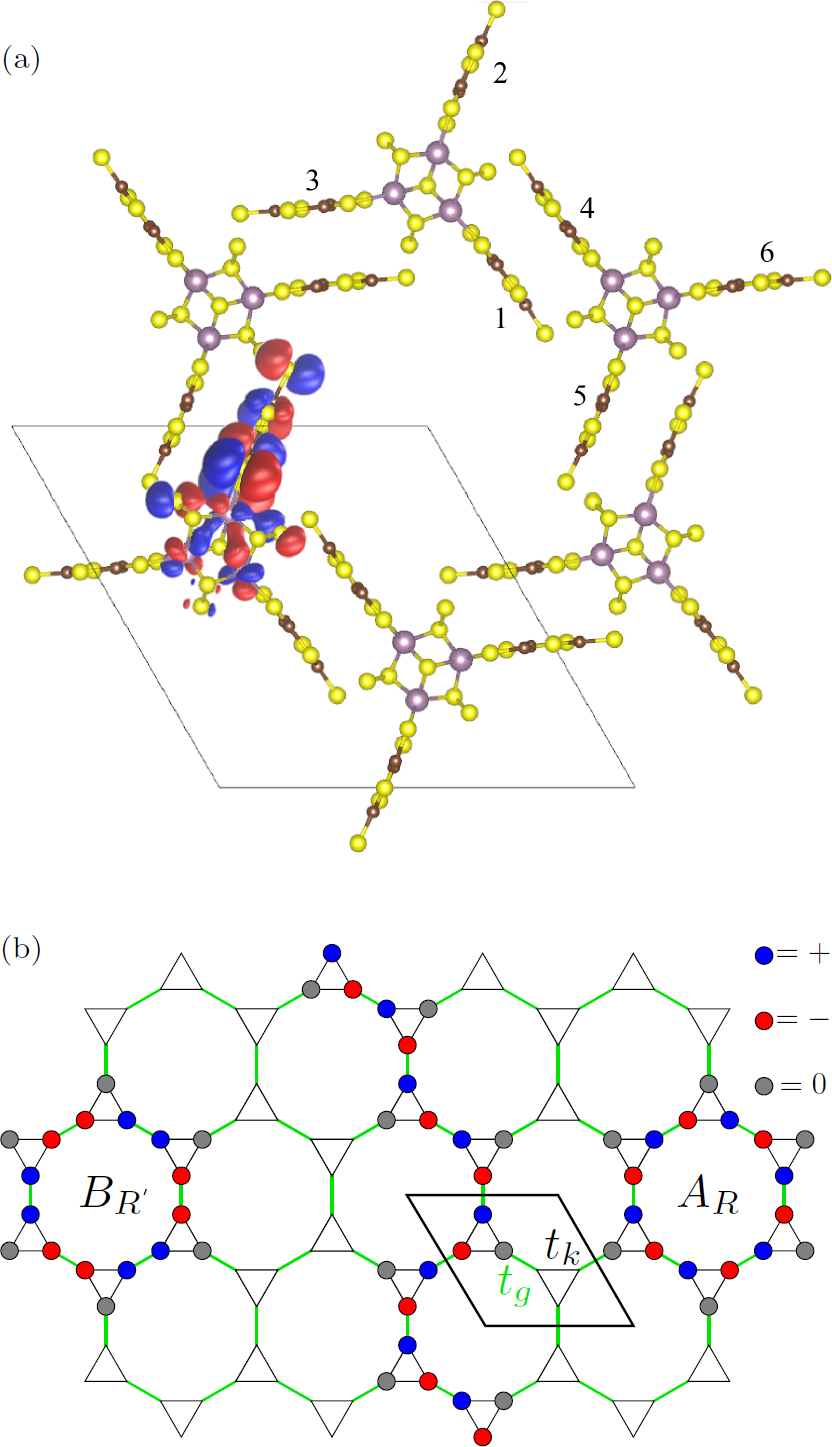} 
\caption{(Color online) (a) Crystal structure of {\moly} projected along the $c$ axis and (b) the kagomene lattice. In (a), one of the Wannier orbitals (WOs) for {\moly} is also shown. Note the predominantly dmit $\pi$-orbital character, with notable Mo$_3$ $d$-orbital character. The other six WOs are related to this one by the $\mathcal{C}_3$ symmetry of the molecules and the inversion symmetry relating the two molecules in each unit cell. Sulfur atoms are shown in yellow, molybdenum in purple, and carbon in brown. In (b) two plaquette states, $A_{\bf R}$ and $B_{{\bf R}'}$, and a topologically non-trivial loop state are shown. 
}\label{fig:structure}
\end{center}
\end{figure}

\moly is a single component organometallic molecular crystal that was first studied as a potential organic metal. Experimentally, one sees an insulator with well-defined localised magnetic moments, but no long-range magnetic order down to the lowest temperatures measured ($T=2$~K, compared to the magnetic interaction $J \sim 100$~K) \cite{llusar04}. Hence this material may have an emergent spin-liquid ground state \cite{janani14a,janani14b}, just one of a few materials have been identified as candidate spin liquids \cite{balents10,lee08,powell11,isono14}. 
Fig. \ref{fig:structure}(a) shows the triangular shape of the molecule and the packing of molecules into a corrugated  
honeycomb lattice with a $P\bar{3}$ ($C_{3i}^1$) space group. The two molecules per unit cell are displaced along the $c$-axis by only slightly less than half the unit cell height, implying a subtle layering of the structure.

In this manuscript, we report \textit{ab initio} electronic structure calculations on {\moly}, which predicts  lines of Dirac fermions and the emergence of 1D behaviour from 3D electronic and crystal structures. We  construct tight-binding models by producing a localized Wannier orbital (WO) representation of the low-energy bands. 
We then focus on the properties of the simplest such model, which allows us to explain the key features of the electronic struture. 

In previous  density functional theory (DFT) calculations \cite{llusar04} antiferromagnetic order was found to give rise to  an insulating state. The band structure predicted for the antiferromagnetic DFT solution is strongly 1D, 
which led to proposals of quasi-1D models for {\moly} \cite{llusar04,janani14a}, and proposals for spin-liquid states on the basis of these models \cite{janani14a,janani14b}.
However, such long range antiferromagnetism has not been observed experimentally; so here we study the paramagnetic solution. Our parameterisation therefore provides a suitable input for quantum many-body calculations that capture correlation effects beyond those included in DFT.
Here we construct a tight-binding model from first principles; we find that the model is fundamentally 3D as the largest in-plane and out-of-plane hopping integrals differ by less than a third. However, we show that interference effects produce two bands that are almost completely flat in-plane, while in the other bands the in-plane bands mass vanishes at the K and K' high-symmetry points. We argue that a Hubbard model at 2/3-filling on this lattice gives rise to a quasi-1D spin-one model. 

We show that both the flat bands (infinite  mass fermions) and massless Dirac fermions occur in a 2D tight-binding model on a honeycomb lattice with each vertex decorated with triangles Fig. \ref{fig:structure}(b). We refer to this lattice as the `kagomene' lattice as it smoothly interpolates between the  kagom\'e and graphene (honeycomb) lattices. Introducing an interlayer hopping between stacked kagomene layers produces the `kagomite' (kagom\'e-graphite) lattice, which displays all the main qualitative features of our DFT band structure. Retaining only the three largest hopping integrals in the Wannier parameterization of the DFT band structure also produces the kagomite lattice.
  
It has recently been shown that the exact ground state of the Kitaev model on the kagomene lattice is a chiral spin liquid \cite{yao07}.
The Kitaev model is a spin-1/2 model, corresponding to $n=3$ electrons per triangular molecule [$n=4$ in {\moly}]. 
It has been shown that, when spin-orbit coupling is included, the kagomene lattice supports topological insulating phases and a quantum spin-Hall effect \cite{ruegg10,wen10}. Topological states have also been predicted on this lattice in the presence of staggered magnetic fluxes \cite{chen12}.

We performed DFT calculations of the electronic structure of {\moly} in an all-electron full-potential local orbital basis using the FPLO package \cite{koepernik99}.
The density was converged on a $(8 \times 8 \times 8)$ $k$ mesh using the PBE generalized gradient approximation \cite{perdew96}.
Localized WOs were constructed from the six  bands closest to the Fermi energy, and with these we computed real-space overlaps to obtain tight-binding parameters \cite{nakamura12,jacko13a,jacko13b,marzari12}. This method has several advantages over the typical procedure of fitting a predetermined tight-binding model to the DFT bands; namely that the parameters relate to real electronic distributions and values of the overlaps are independent of which hopping integrals one includes in the low-energy model. This approach to organic and organo-metallic molecular crystals, and the advantages this technique has in these systems, is discussed in more detail elsewhere \cite{nakamura12,jacko13a,jacko13b}. Most notably, molecular crystals tend to have well-separated bands, so the resulting WO basis is free from the difficulties of disentangling \cite{marzari12}. 

\begin{figure}
\begin{center}
\includegraphics[width=\columnwidth]{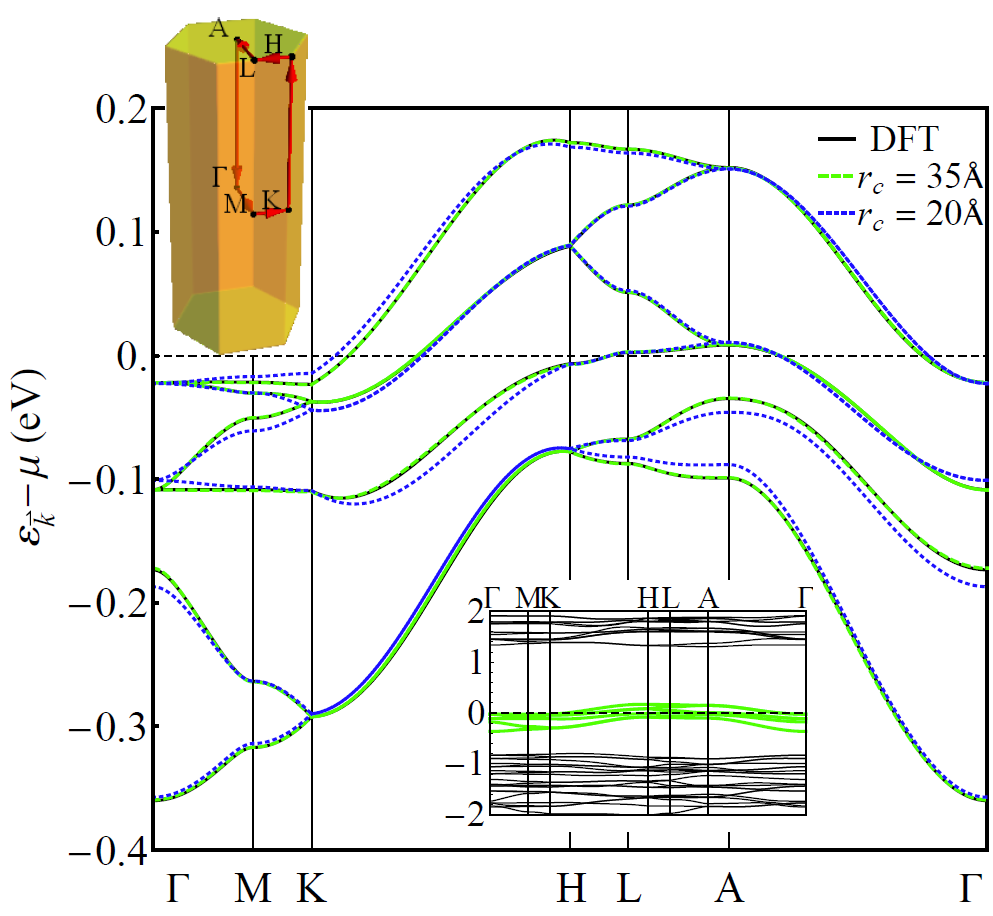}  	
\caption{(Color online.) DFT and tight-binding band structures for {\moly}. The main panel shows the DFT band structure (solid black) along with bands from the tight-binding models constructed from the WOs with a large  ($r_c =  35$~\AA, dashed green) and medium  ($r_c =  20$~
{\AA}, dashed blue)  real space hopping cutoffs. The $r_c =  35$~{\AA} bands lie nearly exactly on top of the DFT bands. 
Lower inset shows the bands in a larger energy window, highlighting the gaps between the frontier bands (green) and the rest of the states (black). Upper inset shows the first Brillouin zone with high symmetry points. }\label{fig:bsdos}
\end{center}
\end{figure}

Fig. \ref{fig:bsdos} shows the DFT band structure of {\moly}. There are six isolated bands around the Fermi energy, all of which have hybrid Mo-dmit character. In the $k_x - k_y$ plane  two pairs of bands meet  at the $K$ and $K'$ points (cf. Fig. \ref{fig:bsdos}, top left inset). Moving along $k_z$, this degeneracy continues along the $K-H$ and $K'-H'$ lines.
Near these lines the dispersion of these bands is linear in-plane, i.e., the in-plane mass vanishes. In the plane the dispersion is highly analogous to graphene, {\it vida infra}, and out of the plane the dispersion is approximately cosine-like.
It is straightforward to show that, as for the honeycomb and  kagom{\'e} lattices  \cite{bernevigbook}, the combination of time-reversal, inversion, and $\mathcal{C}_3$ symmetries pin the Dirac points to the $K-H$ lines. 
Only perturbations that break these symmetries can move the Dirac point away from the $K-H$ line or open a gap.

The two bands that are not involved in the Dirac lines are extremely 1D; with nearly-flat dispersion relations in the $k_z=0$ ($\Gamma-M-K$) plane and cosine-like dispersion in the $k_z$ direction. The bands have slightly larger dispersion in the $k_z=\pi$ ($H-L-A$) plane, but even here the dispersion is very small compared to the in-plane hopping integrals (discussed below).

Fig. \ref{fig:structure}(a) shows one of the WOs for {\moly}. The six orbitals that contribute to the six low-energy bands are all related to this one by the $\mathcal{C}_3$ symmetry of the molecules and inversion symmetry between the two molecules in a unit cell. These six orbitals are the frontier orbitals of the system, and are the basis of our models of its low energy physics.

Note that along the $\Gamma-A$ line the two lower bands are singly-degenerate, and the remaining bands form two doubly degenerate pairs. This is a straightforward consequence of the $\mathcal{C}_3$ symmetry of the  {\moly} molecule. We label creation operators for the three sites of the triangle (WOs on the molecule)  $\hat{c}^\dagger_i, i \in \{1,2,3\}$  (henceforth we  suppress  spin labels). The molecular orbital that belongs to the trivial ($A$) irreducible representation is $\ket{\Psi_A} = \frac{1}{\sqrt{3}}(\hat{c}^\dagger_1 + \hat{c}^\dagger_2 + \hat{c}^\dagger_3)\ket{0}$. The molecular orbitals that belong to the 2D $E$ irreducible representation may be written as $\ket{\Psi_{E{-}}} = \frac{1}{\sqrt{2}}(\hat{c}^\dagger_1 - \hat{c}^\dagger_2)\ket{0}$ and $\ket{\Psi_{E{+}}} = \frac{1}{\sqrt{6}}(-\hat{c}^\dagger_1 - \hat{c}^\dagger_2  + 2 \hat{c}^\dagger_3)\ket{0}$. We stress that, away from the $\Gamma-A$ line, these labels are not good quantum numbers. Nevertheless, we will see below that this basis provides a simple physical description of the band structure.

To produce a low-energy model from the localized WO basis, one must truncate the set of all possible hopping integrals. The simplest method is to include all hopping integrals up to a real space cutoff, $r_c$. 
Fig. \ref{fig:bsdos} compares the DFT band structure with tight-binding models for $r_c = $ 20~{\AA} and 35~{\AA}; the latter is essentially indistinguishable from the DFT bands. All parameters for these models are reported in  the Supplementary Information \cite{SupInfo}. It is worth noting that this model includes significant terms neglected by previous models of {\moly}.

\begin{figure}
\begin{center}
\includegraphics[width=\columnwidth]{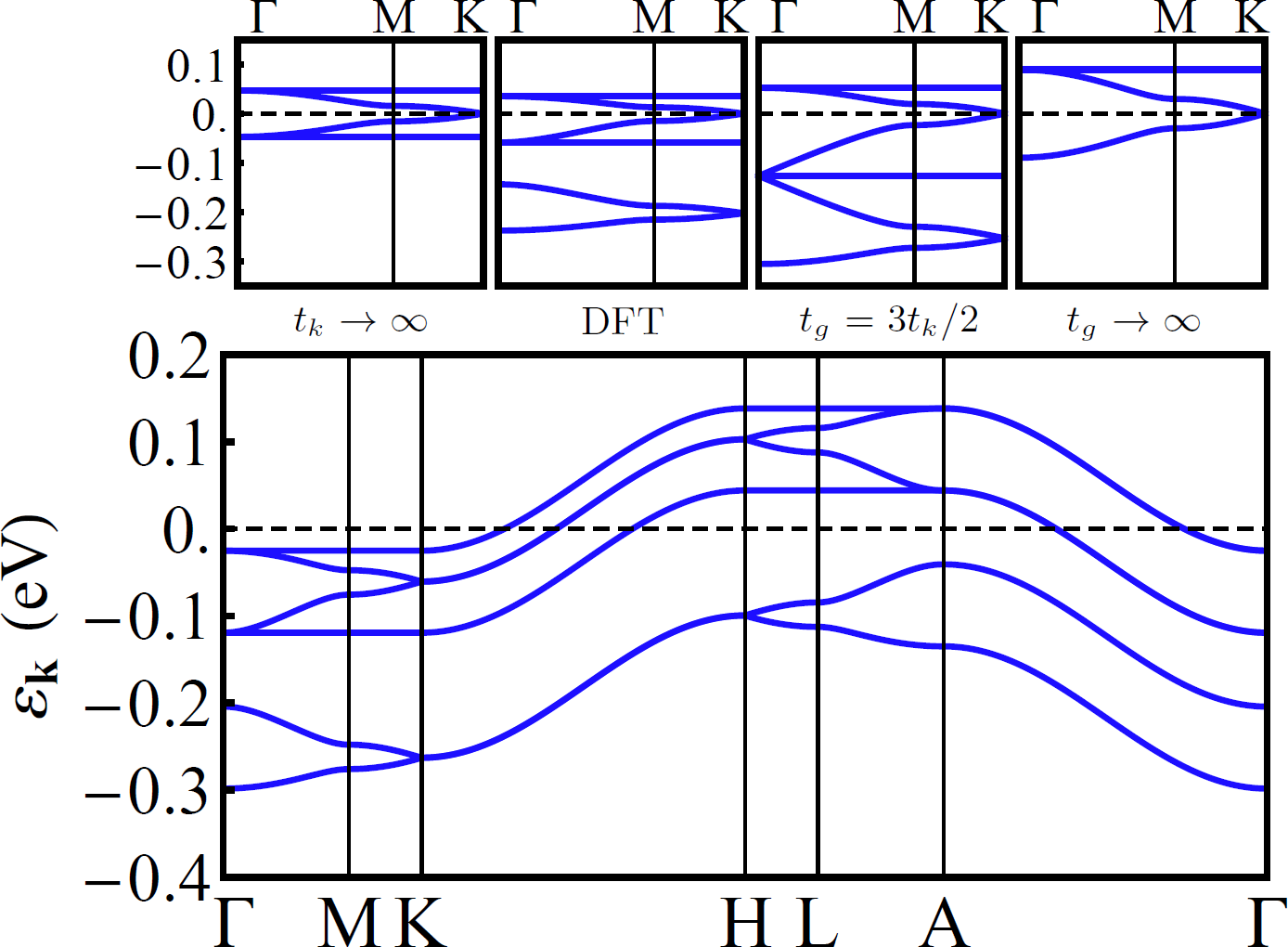}  	
\caption{Main panel: Band structure of the kagomite lattice with $t_k$, $t_g$, and $t_z$ determined from the parameterization of \moly (cf. Table S1 \cite{SupInfo}). Note that the Dirac lines remain and the two `flat' bands are now strictly one dimensional. The insets show the kagomene band structure for, from left to right, $t_k \rightarrow \infty$, the DFT values, $t_g = 3 t_k/2$, and $t_g \rightarrow \infty$ (in each case, the other $t$ takes the DFT values and the filling is fixed at $2/3$). It is clear from the $t_g \rightarrow \infty$ limit that the flat bands on the kagomene lattice are adiabatically connected to the flats bands in the two copies of the kagom\'e lattice \cite{bergman08}.} \label{fig:bskagomene}
\end{center}
\end{figure}

Including only the largest three hopping integrals; $t_k = 59.69$ eV, $t_g = 47.11$ eV, and $t_z = 40.85$ eV (twice as large as any other hopping integral); produces the kagomite model. This model captures all of the main features of the full DFT band structure (compare Figs. \ref{fig:bsdos} and \ref{fig:bskagomene}), including the quasi-1D bands, and lets us explain them. 
This model displays only cosine dispersion in the $k_z$ direction (see Fig. \ref{fig:bskagomene}). 
Thus, all of the non-trivial phenomenology is confined to the plane and below we focus on the 2D  kagomene lattice produced by setting $t_z=0$. 
The origin of the kagomene lattice structure (Fig. \ref{fig:structure}(b)) of \moly can be straightforwardly understood in terms of the crystal structure and WO in Fig. \ref{fig:structure}(a). 
 $t_k$ is the hopping between WOs on the same molecule whereas $t_g$ represents hopping between orbitals on adjacent molecules.

The kagomene lattice interpolates between the honeycomb and kagom{\'e} lattices.
In the limit of $t_g \rightarrow \pm \infty$ WOs on adjacent molecules are strongly hybridized and the model reduces to two decoupled kagom\'e models in the basis of the bonding and anti-bonding hybrid orbitals. Similarly in the molecular ($t_k \rightarrow \pm \infty$) limit, the model is reduced to a honeycomb lattice formed from the $A$ molecular orbitals, another honeycomb lattice formed from the $E{+}$ molecular orbitals, and two flat bands constructed from the $E{-}$ orbitals. 
The transition between the regimes exemplified by these two limits occurs at $t_g = 3t_k/2$ where there is an accidental three-fold degeneracy at $\Gamma$ (see Fig. \ref{fig:bskagomene}). 
Here the lower  flat band switches which of the graphene-like bands it is degenerate with. The DFT bands are in the molecular regime ($t_g < 3 t_k/2$), consistent with intuition.

To understand the flat bands in the kagomene band structure we introduce the  `plaquette states' (sketched in Fig. \ref{fig:structure}(b));
$A_{\bf R}^\dagger = \frac{1}{\sqrt{12}}\sum^{12}_{j=1} \cos{(\pi j )} c_j^\dagger$
and
$B_{\bf R}^\dagger = \frac{1}{\sqrt{6}}\sum^{12}_{j=1}  \cos{(\pi j /2 + \pi/4)} c_j^\dagger$,
where $j$ labels sequential sites around the dodecagon centered at ${\bf R}$, such that sites 1 and 2 are on the same triangle. Explicit calculations shows that both of these states are eigenstates and are localised to a single dodecagon -- this can  be understood as follows.
In both plaquette wavefunctions, amplitudes for the two vertices on the same dodecagon {\it and} the same triangle have the same magnitude and opposite sign, thus it is clear that they are, up to cyclic permutations, $E{-}$ orbitals. 
The hopping terms $t_k\hat{c}^\dagger_3 \hat{c}_1$ and $t_k\hat{c}^\dagger_3 \hat{c}_2$ interfere destructively and cancel exactly. This applies to every triangle around the dodecagon.
The two possible wavefunctions on each ring are either bonding ($B_{\bf R}^\dagger$) or antibonding ($A_{\bf R}^\dagger$) between adjacent triangles (Fig. \ref{fig:structure}(b)); hence they are split by an energy of $2 t_g$ (cf. Fig. \ref{fig:bskagomene}).

These single dodecagon plaquettes can be combined to form extended plaquettes by taking linear combinations of contiguous dodecagon states. For open boundary conditions a sum over all dodecagons produces a plaquette state along the boundary of the system. However, for periodic boundary conditions, this sum vanishes since there is no boundary. Thus for a periodic system of $N$ unit cells, which contains $N$ dodecagons, there are $2(N-1)$ independent plaquette states. However, one can also construct  four more localized states, which are degenerate with the plaquette states, by constructing topologically non-trivial superpositions of $E{-}$ orbitals around the torus, Fig. \ref{fig:structure}(b). There are two distinct non-trivial loops and each may be either bonding or antibonding between triangles.  Because of the non-trivial topology of these states, linear combinations with plaquette states can deform the loops but never close them. Now we have $2(N+1)$ degenerate states whereas each flat band contains $N$ states. These extra states mean that there must be a point at which each of the flat bands touch one of the dispersive bands; this must occur at $\Gamma$ \cite{bergman08}.

Returning to kagomite we see that both the plaquette and topologically non-trival states move with a simple cosine dispersion in the $z$-direction. Additional interactions in the larger $r_c$ models mean that these states pick up a small in-plane dispersion, but it is clear that kagomite provides the basic framework to understand the DFT band structure.

Finally, we turn to the role of interactions in this system. We have not attempted to calculate them directly from our DFT calculations as this is known to be extremely difficult for tightly-packed molecular crystals such as these \cite{scriven09a,scriven09b,cano10,nakamura12}. However, it has recently been shown for Hubbard models on lattices formed of linked triangular molecules, such as the kagomene and kagomite lattices, with $n=4$ electrons per triangle that in the molecular ($t_g\rightarrow0$) limit electronic correlations lead to the formation of localized spin-one moments with Heisenberg interactions \cite{janani14b,janani14a}. In 1D models this insulating phase has been shown to survive even far from the molecular limit \cite{janani14b,janani14a}. 

Each molecule is connected by three $t_z$ hopping integrals to the molecule above/below it, but only one $t_g$ hopping integral connects neighbours in the plane. It is straightforward, but tedious, to show that this means that as $U\rightarrow \infty$, $J_z/J_g \rightarrow  9 (t_z/t_g)^2$. For {\moly}, $t_z^2/t_g^2 = 0.468$, so $J_z \simeq 4.2 J_g$. This suggests that a 1D spin-one model may be appropriate and that Haldane physics may be realised \cite{janani14b,janani14a}, despite the intrinsically 3D crystal structure and hopping integrals.

If interchain interactions are relevant this would lead to a spin-one Heisenberg model on the layered honeycomb lattice. We are not aware of any studies of this model; however, as this lattice is bipartite one might expect that it will display long range order at sufficiently low temperatures if only nearest neighbour exchange interactions are included. However, there has recently been considerable interest in the possibility of spin liquid phases in the spin-1/2 $J_1-J_2$ model on the honeycomb lattice \cite{zhu13,sorella12,meng10}. One is therefore tempted to speculate that second nearest neighbor exchange interactions could lead to a spin liquid state in \moly. Interestingly, the appropriate next-nearest neighbour coupling introduces a chriality into the tight-binding model of \moly, which is not unexpected for the $P\bar{3}$ space group. It is important to note that while field theories of spin liquids are well known, it has proven harder to establish that realistic microscopic Hamiltonians in two or more dimensions have spin liquid ground states \cite{meng10,sorella12,zhu13,yan11,sandvik12,stefan12,balents10,powell11}.

\acknowledgments

We acknowledge discussions with Jaime Merino, Henry Nourse, and Alexander Stilgoe. 
This work was supported by the ARC through grants FT130100161, DP130100757 and LE120100181.

\end{document}


\title{Supplementary Information for: Emergence of quasi-one-dimensional physics in {\moly}, a nearly-isotropic three-dimensional molecular crystal}

\author{A. C. Jacko}\affiliation{School of Mathematics and Physics, The University of Queensland, Brisbane, Queensland, 4072, Australia}
\author{C. Janani}\affiliation{School of Mathematics and Physics, The University of Queensland, Brisbane, Queensland, 4072, Australia}
\author{Klaus Koepernik}\affiliation{IFW Dresden e.V., PO Box 270116, D-01171 Dresden, Germany}
\author{B. J. Powell}\affiliation{School of Mathematics and Physics, The University of Queensland, Brisbane, Queensland, 4072, Australia}

\begin{abstract}
Here we present some supplementary information including an exhaustive list of computed $t$'s for {\moly}, up to $r_c = 35$ {\AA} and $|t| \geq 0.01$ meV.
\end{abstract}

\maketitle

Note that in going from Table \ref{t1} to Table \ref{t2} there is a sharp drop in the magnitudes of the $t$'s, and then they slowly asymptotes to zero (both the sharp drop and asymptote to zero are expected for well localised Wannier orbitals). As one increases the cutoff $r_c$, one includes more and more $t$'s ($N \sim r_c^3$) of lesser and lesser impact on the Hamiltonian. In this system the sharp drop occurs around $r_c = 20${\AA}, hence the use of this cutoff in Fig. 2.

\begin{table}[htbp]
	\centering
		\begin{tabular}{c| r c c}
			\hline  & $t$ (meV) & $|{\bf R}_t|$ (\AA) & ${\bf R}_t$ \\ \hline
			$\mu$ & -50.19 & - &				- \\
			$t_k$ & 59.69 & 11.12 &			${\bf r}_2 -{\bf r}_1$ \\
			$t_g$ & 47.11 & 11.28 &			${\bf r}_4 -{\bf r}_1$ \\ 
			$t_g^{-}$ & 7.40 & 11.81 &	${\bf r}_4 -{\bf r}_1 - {\bf r}_z$ \\
			$t_z$ & 40.85 & 12.39 &			${\bf r}_z$ \\ 
			$t_k^{+}$ & 5.33 & 16.65&		${\bf r}_2 -{\bf r}_1 + {\bf r}_z$ \\
			$t_k^{-}$ & 5.08 & 16.65 &	${\bf r}_2 -{\bf r}_1 - {\bf r}_z$ \\
			$t_h$ & -7.57 & 17.97&	${\bf r}_5 -{\bf r}_1$ \\
			$t_h^{-}$ & 22.88 & 18.32&	${\bf r}_5 -{\bf r}_1 - {\bf r}_z$ \\ \hline
		\end{tabular}
	\caption{List of $t$ parameters, ordered by $|R(t)|$. ${\bf R}_t$ is an example of the path traversed by a hop of magnitude $t$; interactions that are equivalent under $\mathcal{C}_3$ and inversion  are necessarily  the same magnitude. ${\bf r_i}$ labels the origin of the $i$th WO as labeled in Fig. 1(a) of the main text, and coordinates given below. $\bf{r}_z$ is the interlayer lattice vector.}
	\label{t1}
\end{table}

\begin{table}[htbp]
	\centering
		\begin{tabular}{c|c}
			\hline   $t$ (meV) & $|{\bf R}_t|$ (\AA) \\ \hline
  1.12 & 20.54 \\
  1.44 & 21.42 \\
  3.23 & 21.54 \\
  0.72 & 21.82 \\
  0.11 & 24.78 \\
  0.52 & 24.86 \\
  -2.16 & 25.59 \\
  -1.34 & 26.33 \\
  -0.19 & 27.16 \\
  -0.21 & 27.16 \\
  0.29 & 27.54 \\
  -0.03 & 27.69 \\
  1.56 & 27.91 \\
  -1.01 & 27.99 \\
  -1.33 & 28.21 \\
  1.19 & 28.21 \\
  0.34 & 29.10 \\
  -2.45 & 29.10 \\
  -0.10 & 29.53 \\
  0.22 & 32.00 \\
  0.46 & 32.03 \\
  0.17 & 32.03 \\
  0.15 & 32.37 \\
  0.57 & 32.56 \\
  0.93 & 32.58 \\
  0.02 & 32.84 \\
  0.12 & 32.95 \\
  -0.43 & 33.14 \\
  -0.08 & 33.39 \\
  0.01 & 34.57 \\
  0.02 & 34.93 \\		
		\end{tabular}
	\caption{List of $t$ parameters, ordered by $|R(t)|$.}
	\label{t2}
\end{table}

\begin{table*}[ht]
\begin{tabular}{|l|ccc|ccc|c|}
\hline
 & $a$ & $b$ & $c$ & $\alpha$ & $\beta$ & $\gamma$  \\ \hline
{\moly}		&	19.439	&	19.439	&	6.5573	&	90	&	90	&	120		\\ \hline
\end{tabular}
\caption{Lattice parameters for {\moly} from \cite{llusar04}, given in {\AA} and degrees.}
	\label{t3}
\end{table*}

\begin{table*}[ht]
\begin{tabular}{|c|c|}
\hline
Orbital & Coord. (\AA) \\ \hline
WF$_1$ & (0,0,0) \\
WF$_2$ & (7.53648, -8.18075, 0.00000) \\
WF$_3$ & (-3.31650, -10.61716 ,   0.00000) \\
WF$_4$ & (-7.79098, 5.83526, -5.70009) \\
WF$_5$ & (-15.32746,  14.01601, -5.70009) \\
WF$_6$ & (-4.47449 , 16.45241, -5.70009) \\ \hline
\end{tabular}
\caption{Wannier orbital centres for the 6 orbitals per unit cell, given in {\AA}ngstrom, relative to the position of WF$_1$. With these and the unit cell parameters in Table \ref{t3}, one can reproduce the whole network of hopping vectors and make use of the provided $t$'s.}
	\label{t4}
\end{table*}